\documentclass[10pt,preprint]{aastex}

\usepackage{threeparttable}
\usepackage{multirow}
\usepackage{graphicx}

\slugcomment{Not to appear in Nonlearned J., 45.}

\shorttitle{radio fine structures in preflares} \shortauthors{Zhang}

\begin{document}


\title{Solar Radio Bursts with Spectral Fine Structures in Preflares}

\author{Yin Zhang\altaffilmark{1}, Baolin Tan\altaffilmark{1}, Marian Karlick\'y\altaffilmark{2}, Hana M\'{e}sz\'{a}rosov\'{a}\altaffilmark{2}, Jing Huang\altaffilmark{1}, Chengming Tan\altaffilmark{1}, and Paulo Sim\~{o}es\altaffilmark{3}}
\affil{\altaffilmark{1}Key Laboratory of Solar Activity, National
Astronomical Observatories, Chinese Academy of Sciences, Chaoyang
District, Beijing 100012, China}

\affil{\altaffilmark{2}Astronomical Institute of the Academy of
Sciences of the Czech Republic, 25165 Ond\v{r}ejov, Czech Republic}

\affil{\altaffilmark{3}SUPA, School of Physics and Astronomy, University of Glasgow, G12 8QQ, UK}

\email{zhangyin@bao.ac.cn}

\begin{abstract}

A good observation of preflare activities is important for
us to understand the origin and triggering mechanism of solar flares, and
to predict the occurrence of solar flares. This work presents
the characteristics of microwave spectral fine structures as
preflare activities of four solar flares observed by Ond\v{r}ejov
radio spectrograph in the frequency range of 0.8--2.0 GHz. We
found that these microwave bursts which occurred 1--4 minutes before the onset of flares have spectral fine structures
with relatively weak intensities and very short timescales. They include
microwave quasi-periodic pulsations (QPP) with very short period of 0.1-0.3 s and dot bursts with millisecond timescales and narrow frequency bandwidths.
Accompanying these microwave bursts, there are filament
motions, plasma ejection or loop brightening on the EUV imaging
observations and non-thermal hard X-ray emission enhancements
observed by RHESSI. These facts may reveal
certain independent non-thermal energy releasing processes and
particle acceleration before the onset of solar flares. They may
be conducive to understand the nature of solar flares and predict
their occurrence.

\end{abstract}

\section{Introduction}

As early as 1959, Bumba \& K\v{r}ivsk\'{y} found a small flare
(or luminosity increase) preceding a main solar flare and
introduced a term preflare to address the weak effects on
the solar atmosphere produced by them. Early studies have been reviewed by
Martin (1980) and Gaizauskas (1989). Thanks to the development of
recent advanced observing technique for obtaining high spatial and
temporal resolution data, preflare activities are frequently
identified. The common preflare activities include formation of
sigmoid in soft X-ray (Liu et al. 2010), filament activities, such
as rise, oscillation, and eruption (Chifor et. al. 2006; Isobe, \&
Tripathi 2006; Vemareddy et. al. 2012), and transient brightening
in multi-wavelengths (H$\alpha$ (Contarino et al. 2003), UV and
EUV (Cheng et al. 1985), 1600 \AA{} continuum emission (Warren \&
Warshall 2001), soft X-ray (F\'{a}rn\'{i}k et al. 1998), hard
X-ray (Harrison et al. 1985; Tappin 1991; Asai et al. 2005; Chifor
et al. 2007)). These multi-wavelengths studies settle with the
questions concerning the geometry of the magnetic field, the
changing physical properties of the plasma trapped in those
fields, particle accelerations, and the relationship between the
preflare and the main activities.

It is imperative to improve our knowledge of magnetic field and
plasma conditions in current sheets at preflare phase for
resolving the flare problems. Such studies can be guided by radio
observations, which often carry details of the dynamical plasma
processes not visible at other wavelengths. Especially in the
microwave frequency range, it is always regarded as direct signal
of flaring primary energy-releasing and particle accelerations.
Preflare activities in microwave are discovered long ago to
consist of changes in intensity and/or polarization of the
microwaves emitted from an active region some tens of minutes
before the onset of a flare (Kundu, 1965;
Lang, 1980; Hurford \& Zirin, 1982; Kai, Nakajima, and Kosugi,
1983; Xie et al., 1994).
Recently, many fine structures of narrow bandwidth and short
duration which are usually superimposed on smooth background
continuum emission were recorded frequently for the development of
broadband radio spectrometers with high temporal and frequency
resolution (Fu et al. 2004a; Fu et al. 2004b; Huang et al. 2008;
Huang \& Tan 2012; Tan 2013). They are important signatures for
understanding energy release and particle acceleration in solar
flares. Most detailed analyses of microwave fine structures
reported in the literature focus mainly on the impulsive or decay
phase of the flare, with little attention paid to the preflare
phase. Successful observations for doing study on preflare phase
require a high sensitivity and signal to noise ratio of data, and
simultaneous image observations to determine the host region.

The microwave emissions (especially at centimeter and decimeter
wavelengths) originate in the lower corona and the chromosphere,
which provide an independent means of exploring the solar
atmosphere. A good observational understanding of preflare
activities, based on high-sensitivity spectrograms with broadband
in microwave, and high spatial and temporal resolution filtergrams
in multi-wavelengths will certainly help us to understand the
nature of solar flares. After a recent advanced upgrading, the
Ond\v{r}ejov radio spectrographs (ORSC) have the capability to
observe solar microwave emission with high frequency resolution,
high temporal resolution, and high sensitivity. Additionally, with
the launch of Solar Dynamics Observatory (SDO), full disk
filtergrams of the Sun from chromosphere to the corona up to 0.5
$R_{\odot}$ above the solar limb with 0.6 arcsecond spatial
resolution and 12 seconds temporal resolution can be obtained daily
at EUV wavelengths by the Atmospheric Imaging Assembly (AIA, Lemen
et al. 2012). Preflare activities which recorded by ORSC and AIA
simultaneously will provide a unique opportunity to reveal some
key processes of flares, such as the causality consequence, the
actual trigger mechanism and the primary energy release region.
Moreover, despite the differences in size, energy and morphology,
solar eruptions in the solar atmosphere, such as flare, filament
eruption and coronal mass ejection (CME), may be different aspects
of a common physical process involving plasma ejection and
magnetic reconnection (e.g. Shibata, 1999; Priest \& Forbes 2002).
So the preflare activities are important not only for the
initiation of flares, but also for all accompanying eruptive
phenomena. Studying the distinct preflare phenomena could shed new
light on one of the most important puzzle of space weather which
is understanding why flares, filaments and CMEs erupt.

Both from a theoretical point of view (for understanding the flare
phenomenon) and a practical point of view (for forecasting where
and when a flare will occur), it is essential to make a closer
study of preflare conditions in active regions. In this work, we
report the observations of preflare activities in the form of
microwave fine structure of four solar flares. Here, preflare
activity is defined as a transient event preceding the GOES flare at the site of the flare region for which direct
physical associations with flare are implied. The observations and
sample will present in Section 2. The main results will be
addressed in Section 3. Conclusions and discussions will be listed
in Section 4.

\section{Observations and Sample}

\subsection{Observations}


In this work, the preflare microwave fine structures were recorded by ORSC,
which are broadband spectrometers located at Ond\v{r}ejov, the
Czech Republic (Ji\v{r}i\v{c}ka et al. 1993). ORSC includes two
dedicated spectrographs, e.g RT4 and RT5. The frequency range of
RT4 and RT5 is from 2.0 to 5.0 GHz and 0.8 to 2.0 GHz,
respectively. After an advanced upgrading in 2006, ORSC has high
spectral and temporal resolution, and high sensitivity in
broadband microwave frequency range. The spectral resolution is 5
MHz and the temporal resolution is 10 ms. The quick look images
can be obtained from website
(http://www.asu.cas.cz/$\sim$radio/info.htm).

In order to get a comprehensive understanding to the
preflares, we also use other multi-wavelength observations from
several instruments, including:

(1) Phoenix-4 at Bleien Observatory

Since all preflare activities in our sample occurred at the
low frequency side of RT5, we adopt the radio observations in the
low frequency range of 200-800 MHz recorded by a seven meters dish
with a crossed logarithmic periodic antenna (Phoenix-4) at Bleien
observatory (east) to get the full information of the spectral
properties of radio bursts in preflares. Its frequency resolution
is several MHz and temporal resolution is 250 ms. The temporal
resolution of Phoenix-4 is much longer than that of microwave fine
structures. So here, we just use it to show the profile of the
fully radio burst and as a complementary observation to determine
the low frequency limit of some bursts.

(2) the Atmospheric Imaging Assembly on Solar Dynamics Observatory
(AIA/SDO)

AIA/SDO provides multiple simultaneous high resolution full-disk images of the corona and transition region up to 0.5 R$_{\odot}$ above the solar limb with 1.5 arcsecond spatial resolution and 12 seconds temporal resolution (Lemen et al. 2011). Seven narrow EUV bandpasses centered on specific lines: Fe XVIII (94 \AA{}), Fe VIII, XXI (131 \AA{}), Fe IX (171 \AA{}), Fe XII, XIIV (193 \AA{}), Fe XIV (211 \AA{}), He II (304 \AA{}), and Fe XVI (335 \AA{}) have been employed. The temperature diagnostics of the EUV emissions cover the range from $6\times10^{4}$ K to $2\times10^{7}$ K (Lemen et al. 2011). It provides the topological evolution information in the related source regions associated to the preflare activities. We choose four wavelengths of 1600 {\AA},
304 {\AA}, 171 {\AA} and 193 {\AA}, which contain the emission
from the solar photosphere, chromosphere, corona and hot flare
plasma.

(3) Nancay Radioheliograph (NRH)

NRH is operated by the Observatoire de Paris at 10 frequencies
between 150 and 450 MHz. It consists of 44 antennas of size
ranging from 2-10m, spread over two arms (EW and NS) with
respective lengths of 3200 m and 2440 m. The resolution of the 2
dimensional (2D) images depends on the frequency and the season.
Roughly speaking, the resolution is approximately 5.5$^{'}$ - 3.2$^{'}$ (164 MHz) and
2.2$^{'}$- 1.25$^{'}$ (432 MHz). NRH observations will help to
determine the source region of the radio emission.

(4) The Reuven Ramaty High Energy Solar Spectroscopic Imager
(RHESSI)

RHESSI is a NASA Small Explorer Mission, launched on February 5,
2002. It images solar flares from soft X rays (3 keV) to gamma
rays (up to 17 MeV) and provides high resolution spectroscopy up to
gamma-ray energies of 17 MeV. Furthermore, it has the capability
to perform spatially resolved spectroscopy with high spectral
resolution. From RHESSI observation, we may get information about
the non-thermal processes associated to the preflares. (Lin et al.
2002).

(5) Soft X-ray (SXR) telescope on Geostationary Operational
Environmental Satellites (GOES)

SXR/GOES provides continuous monitoring of full-disk solar SXR
intensity at 0.5-4 \AA~ and 1-8 \AA~ channel with a minimum
cadence of 2 s. It is the most important indicator to show the
whole processes of solar flares.

\subsection{Sample}

The sample of radio bursts which we studied in this work is during the time period from 2010 April (after the launch of SDO) to 2013 June. During this time period, 2996 flares (including C, M, and X class flares) were recorded by GOES and 240 radio bursts were recorded by ORSC. Out of the 240 radio bursts, 156 radio bursts were associated with GOES flares. Out of the 156 flares associated radio bursts, 18 radio bursts started prior to the associated GOES flares. In order to determine the spatial correlation between radio bursts and flares without corresponding image observations, we delete events which occurred on the solar surface with numerous active regions at first. Then, 14 events remained. 
For the 14 events, the following steps were used to confirm the spatial correlation between radio bursts and the
corresponding flare events: (1) Radio images of Nancay
Radioheliograph (NRH) were used if preflare radio bursts occurred at NRH observational frequencies. Even though the spatial resolution of NRH is about several arcminutes, it is enough to determine whether radio burst has the same host region with the flare or not; (2) the flare host region is an isolated active region on the solar disk and there are no other activities which can be identified from SDO observations at wavelengths of 304 {\AA} and 193 \AA. Finally, four flares remained. We show the event list in Table 1. Out of the four events, three events (C3.2 flare event on 2010 August 14, C5.5 flare event on 2011 December 25, and M1.9 flare on 2012 May 07) can be confirmed by both criteria. And the C4.4 flare on 2010 August 01 is confirmed by SDO data.

\begin{deluxetable}{ccccccccccccccccccc}
\tablecolumns{16} \tabletypesize{\scriptsize} \tablewidth{0pc}
\tablecaption{The List of the Preflare Events with Microwave
Spectral Fine Structures\label{tbl-1}} \tablehead{
 \colhead{    }  &\colhead{        }&\colhead{     } & \colhead{GOES}  & \colhead{Flare}& \colhead{ }   && \colhead{ }    & \colhead{ }     & \colhead{Radio}           & \colhead{burst}     & \colhead{ }        & \colhead{ } && \colhead{HXR}   & \colhead{ }   \\
 \cline{3-6} \cline{8-13} \\
 \colhead{Date}  &\colhead{Position}&\colhead{class} & \colhead{start}& \colhead{peak} & \colhead{End} && \colhead{Type} & \colhead{start} & \colhead{$\Delta$t} & \colhead{f$_{high}$}& \colhead{f$_{low}$}& \colhead{P} && \colhead{12--25} & \colhead{EUV}   \\
 \colhead{}      & \colhead{    }   & \colhead{}     & \colhead{(UT)} & \colhead{(UT)} & \colhead{(UT)}&& \colhead{    } & \colhead{(UT)}  & \colhead{(s)}       & \colhead{(GHz)}       & \colhead{(GHz)}      &\colhead{(s)}&& \colhead{(KeV)}   & \colhead{ }     \\}
\startdata
 2010-08-01      & N20E36           &    C3.2        &07:56 &08:26&09:46&&dot &  07:53:50       &           &    0.9      &   $<$0.8       &            &&     I           & LB    \\
 2010-08-14      & N17W52           &    C4.4        &09:41 &09:59&11:10&&dot &  09:37:12       &     40    &   0.9       &    0.5         &            &&     I           & FE,LB    \\
                 &                  &                &      &      &    &&dot &  09:40:00       &           &    $>$2.0   &    0.35        &            &&     I           &    \\
 2011-12-25      & S32W16           &    C5.5        &08:49 &08:55&09:01&&QPP &  08:47:16       &     3.5   &     1.35    &    1.035       &     0.2    &&     I           & FE,LB    \\
                 &                  &                &      &     &     &&QPP &  08:47:20       &     2.3   &     1.00    &    0.6         &     0.2    &&     I           &   \\
                 &                  &                &      &     &     &&QPP &  08:47:26       &     0.5   &     1.00    &    0.6         &     0.1    &&     I           &     \\
                 &                  &                &      &     &     &&QPP &  08:47:32       &     1.2   &     1.00    &    $<$0.8      &     0.1    &&     I           &    \\
                 &                  &                &      &     &     &&QPP &  08:47:36       &     1.4   &     1.10    &    $<$0.8      &     0.3    &&     I           &    \\
                 &                  &                &      &     &     &&QPP &  08:47:39       &     0.5   &     1.00    &    $<$0.72     &     0.1    &&     I           &      \\
                 &                  &                &      &     &     &&QPP &  08:48:35       &           &     1.05    &    $<$0.8      &     0.3    &&     I           &     \\
 2012-05-07      & S19W46           &    M1.9        &14:03 &14:31&14:52&&dot &  14:02:09       &     10.4  &     1.45    &    0.55        &            &&    $\diagup$    &  FE,LB   \\
                 &                  &                &      &     &     &&dot &  14:02:24       &     18.4  &     1.45    &    0.55        &            &&  $\diagup$     &    \\
\enddata

\tablecomments{$\Delta$t: duration of the burst group; $\diagup$:
no RHESSI data; I: RHESSI flux (12--25 keV energy band) increases associated with the preflare radio
bursts; FE: filament eruption; LB: loop brightening}
\end{deluxetable}

\section{Analysis Results}

\subsection{The Properties of the Preflare in a C5.5 Event on 2011 December 25}

\begin{figure}
\centering
\includegraphics[width=12 cm]{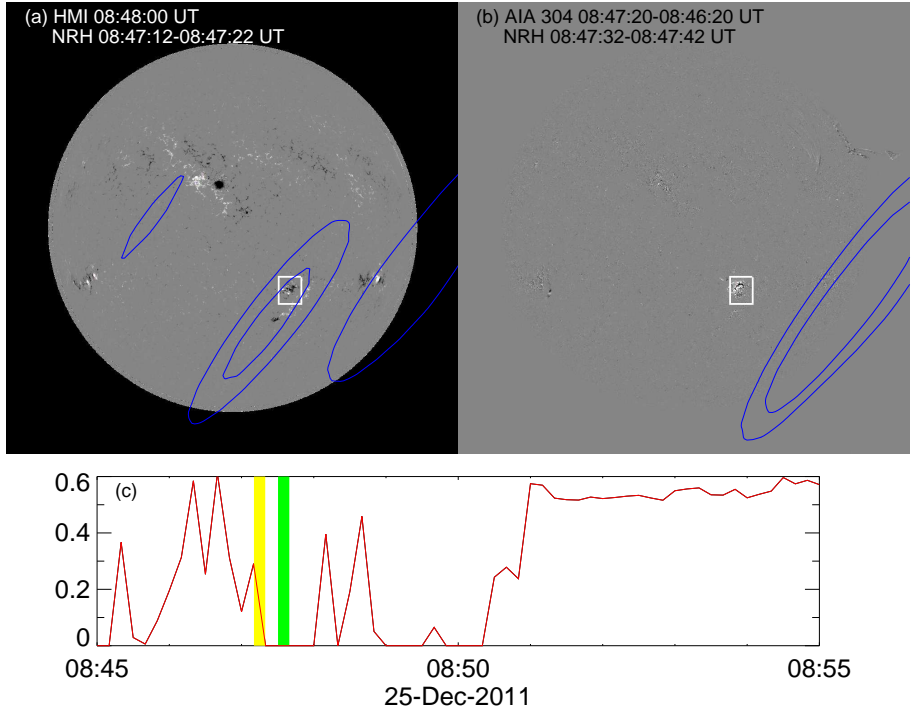}
\caption{(a) line of sight magnetograph observed by HMI with NRH contours (228 MHz); (b) running difference image of 304 {\AA} observed by AIA with NRH contours (228 MHz); (c) time profile of the radio flux in the flare host region at 228 MHz observed by NRH. The yellow and green shadow outlines the integration time of the NRH contours in panel a and b.}
\label{Fig:fig_hmi_nrh}
\end{figure}

On 2011 December 25, a C5.5 flare occurred in a newly emerging
active region NOAA 11387 (S32W16). According to the GOES record,
the flare started at 08:49 UT, peaked at 08:55 UT and ended at
09:01 UT. It is an impulsive flare with the impulsive phase lasting for about 6 minutes. According to ORSC record, the main radio burst which associated with the GOES flare appears around 08:48:35 UT. Meanwhile, there are some impulsive radio bursts with relatively weak intensity and very short duration occurred prior to the main radio burst.
Fig. \ref{Fig:fig_hmi_nrh}a is the line of sight magnetic field observed by HMI, with the flare host region outlined by white square. It shows that the active regions on solar surface at that time are few and far apart. Fig. \ref{Fig:fig_hmi_nrh}b is a running difference image of 304 {\AA} observed by AIA at 08:47:20 UT. Its base image is taken at 08:46:40 UT. It shows that the initial brighting, which occurred in the flare host region and temporal correlated with the microwave burst, is the unique activity on the solar surface at that time. This event was observed by NRH at several frequencies. Blue contours in Fig. \ref{Fig:fig_hmi_nrh}a and b show the source regions of the microwave observed by NRH at 228 MHz. They are set at 50\% and 80\% of the maximum brightness. The NRH contours shown in Fig. \ref{Fig:fig_hmi_nrh} were integrated during the time period from 08:47:12 to 08:47:22 UT (Fig. \ref{Fig:fig_hmi_nrh}a) and from 08:47:32 to 08:47:42 UT (Fig. \ref{Fig:fig_hmi_nrh}b). These time period were also outlined by yellow shadow and green shadow in Fig. \ref{Fig:fig_hmi_nrh}c, respectively. Contours in Fig. \ref{Fig:fig_hmi_nrh}b show a unique source region which is persistent and slowly varying during NRH observations. And even during the flaring phase, this source region can be identified from time to time. Contours in Fig. \ref{Fig:fig_hmi_nrh}a show two additional transient source regions. The one located on the south side is co-spatial with flare host region. The time profile of the flare host region recorded by NRH at 228 MHz was shown in Fig. \ref{Fig:fig_hmi_nrh}c. In plotting this time profile, only pixels whose flux is greater than 50\% of the maximum are counted. This time profile shows that the emission from the source region changes very rapidly during the preflare phase and then maintains at a high level during the flare. It is known that active regions can emit radio signals without a flare at 228 MHz, possibly through the thermal bremsstrahlung mechanism, but it varies very slowly. So we suggest that the rapidly varying emission which emit from the flare source region is related with the preflare activities. 

\subsubsection{Microwave Spectral Structures in the Preflare}

\begin{figure}
\epsscale{.8} \plotone{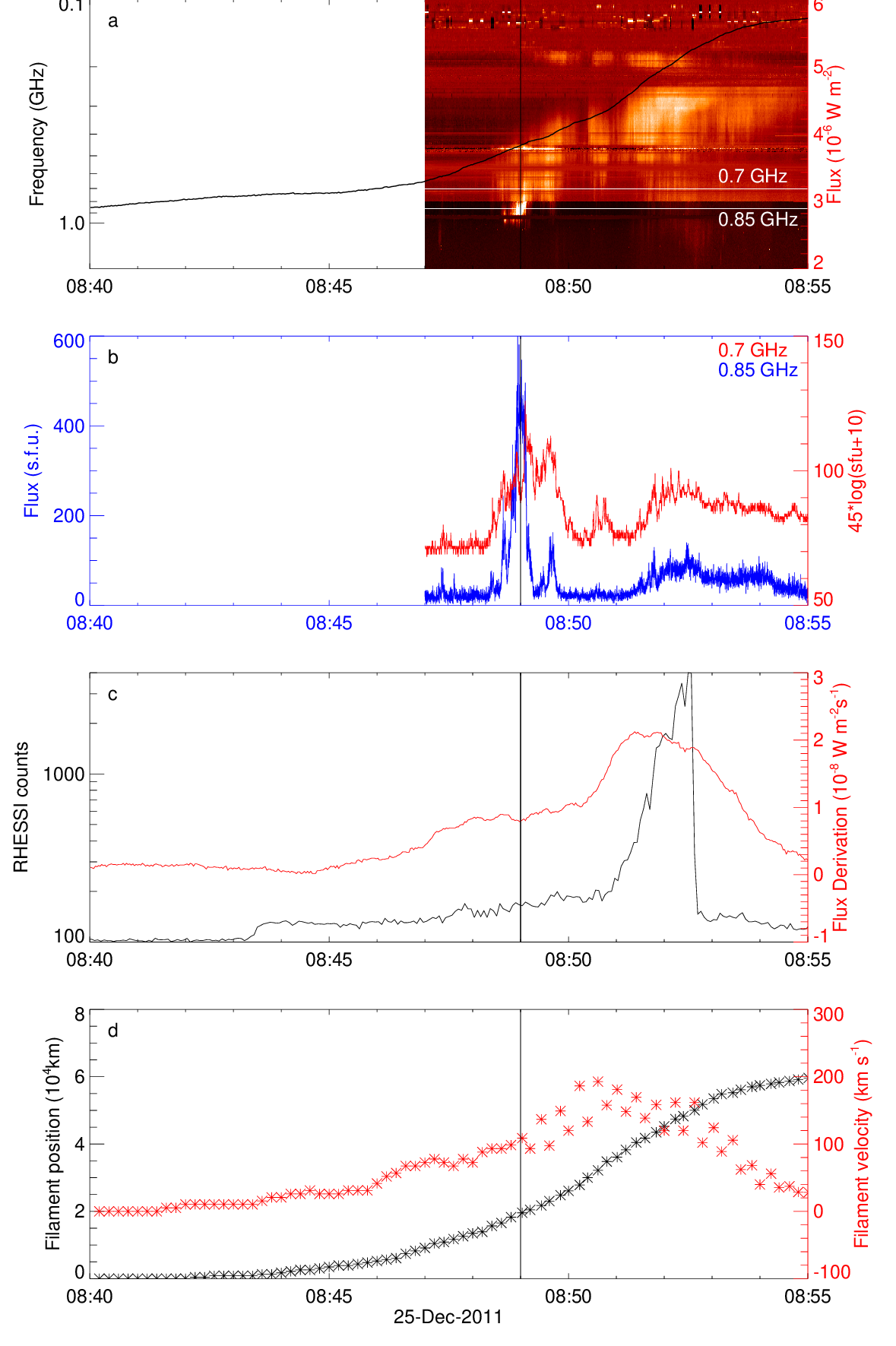}
\caption{(a) The dynamic radio spectrum of 2011 December 25 event at
0.2--1.6 GHz; the vertical lines in all panel represent the start of
the GOES flare;(b) Temporal profiles of radio burst at 0.7 and 0.85
GHz; (c) Red line: Time derivative of GOES X-ray intensity profile
at wavelength of 1--8 \AA; Black line: RHESSI time profile at 12--25
KeV energy band; (d) Temporal profile of filament displacements
(black stars) and velocity (red stars).}
\label{Fig:fig01_spectrum_flux}
\end{figure}

Fig.
\ref{Fig:fig01_spectrum_flux}a shows the dynamic microwave spectrum during the
preflare and impulsive phase in the frequency range from 0.2 to
1.4 GHz. The black vertical line indicates the starting time of
the flare. The microwave spectrum shows that several
independent radio bursts appear about 2 minutes before the
GOES flare, eg. around 08:47 UT. They have relatively weak
intensity with very short durations (several second). And around
08:48 UT, less than 1 minute before the flare, the main
radio burst begins to appear. The main burst is a slow drifting
structure, with long duration (more than six minutes) and strong
intensity. The spectrum shows that the preflare microwave bursts
are in the form of individual bursts. Meanwhile the early phase of the
main burst starts prior to the flare and lasts during
the whole impulsive phase.

The temporal profile of the radio bursts at 0.85 GHz recorded by
ORSC and at 0.7 GHz recorded by Phoenix-4 are shown in Fig.
\ref{Fig:fig01_spectrum_flux}b.
The former profile shows abundant of
substructures while the latter not. Moreover, the flux of the preflare is only about 7\% of the maximum intensity of the main burst.

\begin{figure}
\epsscale{.8} \plotone{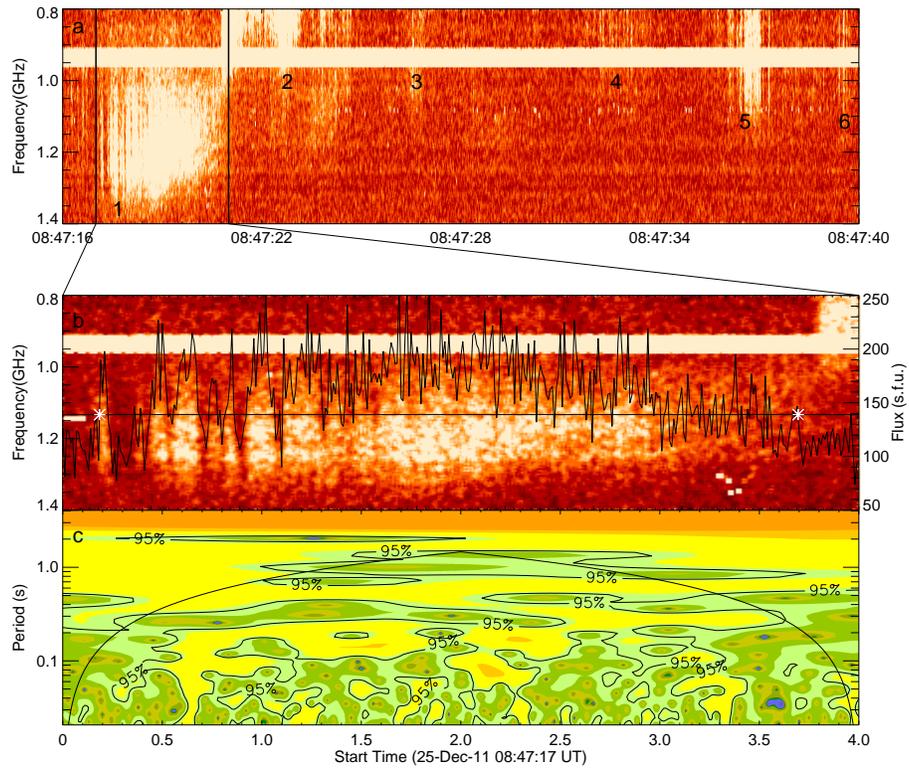}
\caption{Radio spectrogram of the drifting quasi-periodic
pulsation and related wavelet power spectra at 1.05 GHz.}
\label{Fig:fig02_wavelet}
\end{figure}

The detailed dynamic spectrum of the independent radio bursts
recorded by ORSC during the period from 08:47:16 to 08:47:39 UT
are expanded in Fig.
\ref{Fig:fig02_wavelet}a. From the spectrum, six groups of
drifting structures are identified. The frequency bandwidth of the
bursts were calculated by following steps. Firstly, we integrated
the radio burst through its duration. The background flux is
defined as the sum of the mean value of the quiet sun and 3 times
of its standard deviation ($3\sigma$) at the corresponding
frequency. Then the start and end frequency is defined as the
frequency at which the flux is greater than the background flux.
Meanwhile, the duration is defined as the time interval that the
radio flux greater than the background flux. The middle panel
shows the first group burst with fully temporal resolution. The
radio flux at 1.2 GHz were overlapped on it. The transverse line
indicates the background flux at that frequency. Stars represent
the start and the end of the burst. It shows that the first burst
group occurred around 08:47:16 UT, with duration of about 3.5 s
and the frequency range from 1.035 GHz to 1.35 GHz. The properties
of other burst groups are listed in Table 1.

Meanwhile, Fig.
\ref{Fig:fig02_wavelet}b shows that the radio burst is composed of
many drifting pulsations. These pulsations scatter over the image
along the temporal axis with different time interval at the early
stage and then overlap each other. Careful examination of the first
few pulses revealed that they first occur at low frequency and
then shift to the high frequency rapidly. The magnitude of the
frequency drifting rate is about -10 GHz\,s$^{-1}$. For a single
pulse, the duration is about several tens of milliseconds and the bandwidth is about 0.3 GHz. To study the periodic property of the pulsating structure, we adopt
wavelet analysis on the observational data. The wavelet power
spectrum shown in Fig.
\ref{Fig:fig02_wavelet}c was computed from time series at
frequency of 1.15 GHz. The solid black contour is the 95\%
confidence level for the red noise. Areas outside the contour are
regions of the wavelet spectrum where edge effects become
important, the so called cone of influence. The wavelet power
spectra shows that there is a strong periodic component at 0.2 s.
This periodicity is relatively stable. There is another periodic
component within 95\% confidence level which occurs at 0.4 s. After two seconds later, these two periodic components
shift slightly from a period near 0.4 s to a period closer to
0.7 s, and from a period near 0.4 s to a period closer to 0.3 s,
respectively. The wavelet power spectra confirms that the
pulsating structures are quasi-periodic pulsations (QPP). There is
no obvious global frequency drifting when we considered the burst as a whole
pulsation structure.

Table\,1 shows the main properties of the six QPPs occurred in the
preflare. In this table, we do not give the duration time of radio
burst which occurred prior to the flare and lasted during the
impulsive phase. From Table\,1, we find that all the preflare QPPs
are occurred in the frequency range from 0.6 to 1.3 GHz, the
durations are in the time period from 0.5 to 3.5 s. Their periods are in
the range of 0.1 -- 0.3 s, which belong to very short period
pulsation (Tan et al. 2010).

\subsubsection{The Properties of the Source Region in the Preflare}

\begin{figure}
\centering
\includegraphics[width=9 cm]{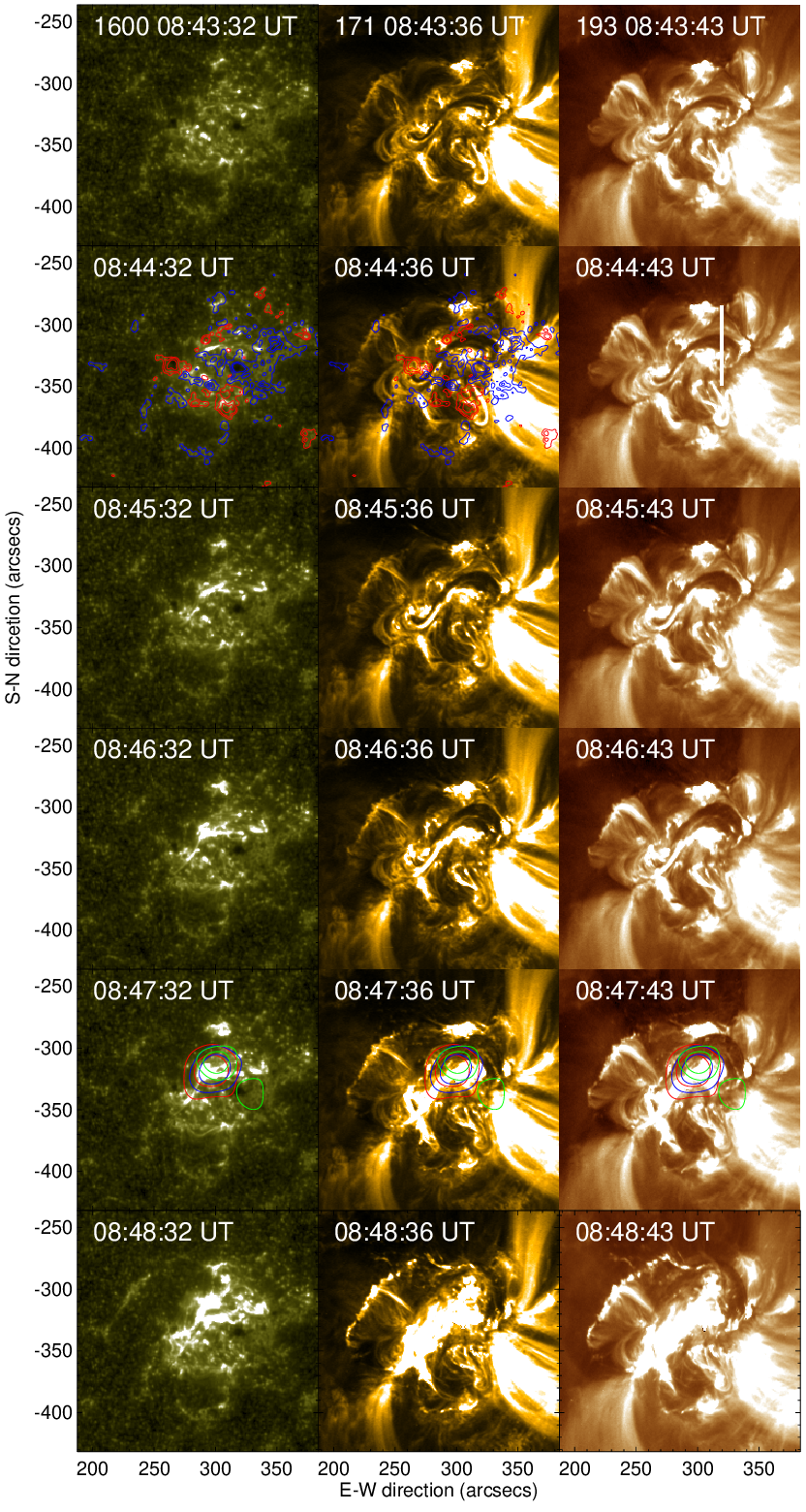}
\caption{Dynamic activities observed by AIA/SDO at wavelength of
1600 \AA{}, 171 \AA{}, and 193 \AA{} in the preflare phase of the
event on 2011 December 25. Magnetic field observed by HMI/SDO was
overlaid on 1600\AA{} and 171 \AA{} images at 08:44:36 UT. Snapshots
of AIA 171 \AA{} and 193 \AA{} around 08:47 UT were overlaid by
contours indicating RHESSI emission at 3 -- 6 keV (red), 6 -- 12 keV
(blue), and 12 -- 25 keV (green). Contour levels are set at 60\% and
80\% of the maximum brightness of each individual image. Vertical
line in 193 \AA{} image at 08:44:43 UT represents the slice, along
which the displacement of the filament were calculated.}
\label{Fig:fig03_sdo_evolution}
\end{figure}

In order to get the information of the source region in the
preflare, here we adopt the multi-wavelength observations. Fig.
\ref{Fig:fig03_sdo_evolution}
shows the dynamic evolution of the preflare activities in the UV
and EUV imaging observations recorded by AIA/SDO. Magnetic field
obtained by Helioseismic and Magnetic Imager on SDO (HMI/SDO) was
overlaid on 1600 \AA~ and 171 \AA~ images around 08:44 UT, with
blue for negative polarities and red for positive polarities. The
most distinctive feature of the host region is that there is an
S-shaped filament (positive magnetic helicity) crossing over the
neutral line, with the ends near (260, -340) and (310, -330) in
the solar disk coordinate system. The presence of an S-shape
filament is regarded as the evidence for current-carrying twisted
or sheared magnetic fields which possess magnetic helicity. Active
regions with this kind of morphology are more likely to erupt than
others (Canfield, et al., 1999).

The beginning of the eruption is marked by a very gradual but
noticeable rise of the right part of the S-shape filament as
early as 08:42 UT. The displacement of the filament was traced
along a slice as indicated by a white vertical line in 193 \AA~
image at 08:44:43 UT. A time-distance diagram of the filament
along the slice was shown in Fig.
\ref{Fig:fig01_spectrum_flux}c. It shows that the
slow-rising of the filament can be best observed between 08:42 - 08:47 UT with a
steady upward motion of the filament. Its speed is around
20\,km\, s$^{-1}$. The initial intensity enhancements of the flare
can be found easily from 1600 \AA{} image, which is situated
under the middle of the filament around 08:44:43 UT. The EUV
brightening propagated in the direction along the polarity inverse
line from the initial brightening area, and extended away from the
polarity inverse line. The development of two flare ribbons can be
seen clearly in 1600 \AA{} images. Brightenings extended in 171
\AA~ and 193 \AA~ images associated with the two ribbons are the
dynamic evolution loops in low atmosphere which locate under the filament.
Associated with the flaring process, the microwave bursts occurred
around 08:47 UT. The filament speed was found to increase around
the same time, and to
reach a maximum of 200\,km\, s$^{-1}$ around 08:51 UT.

\begin{figure}
\centering
\includegraphics[width=9 cm,angle=-90]{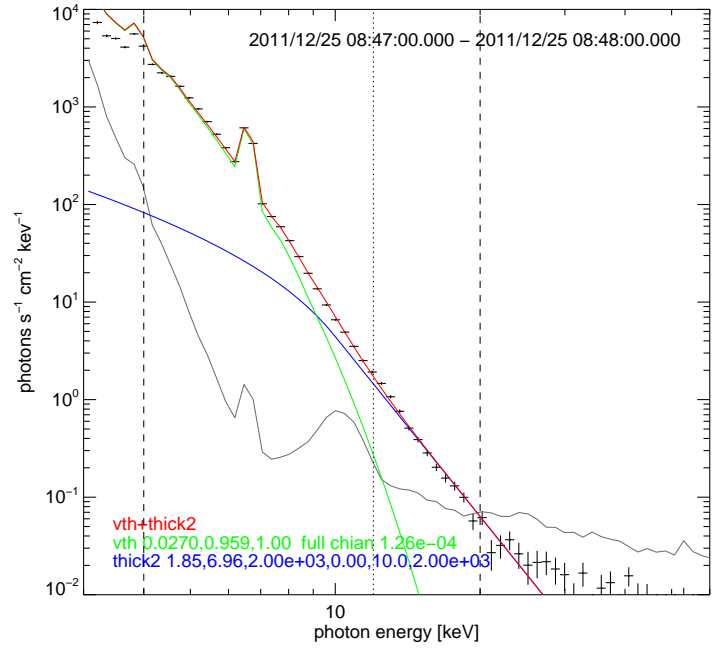}
\caption{RHESSI spectrum (plus signs) accumulated between
08:47 and 08:48 UT. The thermal spectrum
is fitted with the variable thermal component (green curve), while the non-thermal portion is fitted with a single power law (blue curve). The sum is represented
by the red curve. The fitting range used was 3--20 keV.}
\label{Fig:rhessi_spectrum_20111225_0847}
\end{figure}

We analysed the X-ray spectrum of the preflare brightening
observed with RHESSI by means of imaging spectroscopy with the
Object Spectral Executive (OSPEX) (Dere et al. 1997; Landi et al.
2006). Fig.
\ref{Fig:rhessi_spectrum_20111225_0847} shows the spectrum of thermal and non-thermal X-ray
emission associated with the individual bursts during the period
from 08:47 to 08:48 UT. Thermal (low energy) portion is fitted
with the variable thermal component while the non-thermal (high
energy) portion is fitted with a single power law. Fig.
\ref{Fig:rhessi_spectrum_20111225_0847} shows
that at energies above $\sim$9 Kev, the non-thermal contribution
dominates over the thermal component. The temporal profile of
non-thermal emission which recorded by RHESSI at 12--25 KeV energy
band, is shown in Fig. 1c. Meanwhile, the red curve shows the time
derivative of the GOES 1--8 \AA{} flux. Both curves show obviously
increase right after the first group of microwave burst. The
RHESSI flux increase in the preflare phase is small, so the
integration time of the flux curve which is shown in Fig.
\ref{Fig:fig01_spectrum_flux}d is 4
seconds. It is too long to find a counterpart of the QPP in RHESSI
observation.  However, the microwave QPPs, which occurred just
before the maximum of an X6.9 class flare, accompanying hard X-ray
QPPs were studied by Tan \& Tan (2012).

The X-ray brightening, between 08:47 -- 08:48 UT, observed at
energy of 3 -- 6 keV (red), 6 -- 12 keV (blue), and 12 -- 25 keV
(green) are overlaid on 304 \AA~, 171 \AA~ and 193 \AA~ image around 08:47
UT. Contour levels are set at 60\% and 80\% of the maximum
brightness of each energy band. It reveals the presence of single
source in 3 -- 6 keV and 6 -- 12 keV. Single source at difference energy band was
found location on the polarity inversion line and spatially
coincident with the initial brightening. The HXR emission in the
energy band of 12 -- 25 keV has two sources. One is spatially
consistent with the low energy sources, which covers the initial
brightening at the preflare phase and locates on the top of the
two ribbons. The figure indicates that the maximum of X-rays sources
at softer energies (3 -- 6 keV and 6 -- 12 keV) and harder energy
(12 -- 25 keV) are at the top of loops (that may be formed below
rising filament). The other source at 12--25 KeV energy band
encircle the foot of other loops. While no intensity enhancement
at EUV wavelengths (1600 \AA, 171 \AA, and 193 \AA) was
identified spatial corrected with this source. These X-ray
sources remained at the same place through the whole flaring
process.

Briefly, in the preflare phase of the C5.5 event on 2011 December 25,
microwave QPPs with period of subsecond appeared in the frequency
range of 0.6 GHz - 1.35 GHz. Accompanying these microwave QPPs,
nonthermal hard X ray emission and the motion of a sigmoid
filament is observed by multiple wavelength imaging
observations.

\subsection{The Properties of Preflares in Other Three Events}

Here, we present other three flare events which are also
discerned to have microwave spectral fine structures in their
preflare phases.

\subsubsection{C3.2 Event on 2010 August 01}

\begin{figure}
\epsscale{.8} \plotone{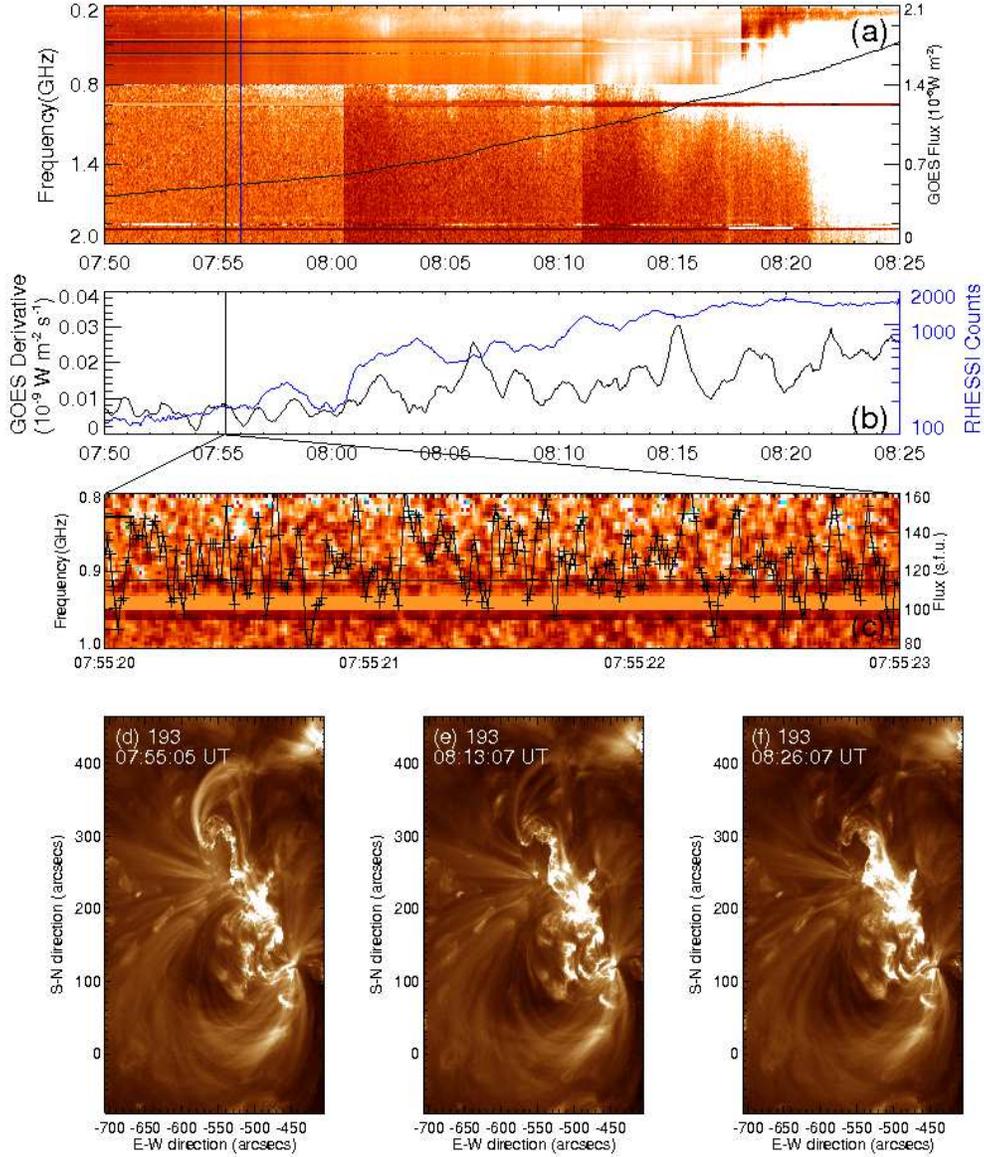}
\caption{The flare event on
2010 August 01. (a) The dynamic radio spectrum. Black curve is the
GOES flux in 1--8 \AA{} energy band. Blue vertical line indicates
the start of the GOES flare; (b) The blue curve is the RHESSI counts
at the energy band of 12--25 Kev and the black curve is the
derivative of the GOES soft X-ray flux in 1--8 \AA{} energy band;
(c) The dynamic radio spectrum with fully temporal resolution during
the period which indicates by black vertical line in panel (a)
and (b); (c)--(e) Dynamic evolution of the event observed by AIA.}
\label{Fig:fig_20100801}
\end{figure}

The C3.2 flare on 2010 August 01 occurred on the heliographic
location 20N 36E. The flare started at 07:56 UT, peaked at 08:26
UT and ended at 09:46 UT. It is a long duration flare with the
impulsive phase lasting for 30 minutes. The Fig.
\ref{Fig:fig_20100801}a shows the dynamic microwave spectrum
during its preflare and impulsive phase in the frequency range
from 0.2 GHz to 2.0 GHz. Since the microwave flux varies greatly
during the flare period and the sensitivity of two instruments is
different, the spectrum is shown in different contrast to show the
fine structures of the microwave bursts in difference phase. The
overlapped black curve is GOES flux profile at 1--8 \AA. It shows
that GOES flux increases very slowly. The blue vertical line
indicates the start of the GOES flare. The spectrum shows that
radio burst appeared around 07:53:30 UT at frequency of about 0.3
GHz observed by Phoenix-4 and 0.8 GHz observed by ORSC. The radio
bursts in Phoenix-4 is a narrow band burst, which extends to the
main burst. While the radio burst in ORSC first appears at the low
frequency side and is confined in a narrow frequency band for
several minutes. Its counterpart in the lower frequency can not be
identified in Phoenix-4, because the different sensitivity between
the two instruments. Associated with the development of the flare,
the intensity of the radio burst increases and the frequency band
extends.

The simultaneous hard X-ray flux recorded by RHESSI at 12--25 KeV
energy band was shown in Fig.\,\ref{Fig:fig_20100801}\,b with blue
curve. It shows that the hard X-ray flux begins to increase from
07:55 UT. The black curve shows the time derivative of the GOES
1--8 \AA{} flux.

The fully temporal resolution dynamic spectrum during the time
period from 07:55:20 to 07:55:23 is shown in
Fig.\,\ref{Fig:fig_20100801}\,c. The vertical black line in
Fig.\,\ref{Fig:fig_20100801}\,a and b indicates the time point
which is shown in Fig.\,\ref{Fig:fig_20100801}\,c. Radio flux at
0.83 GHz is overlapped on the spectrum and the horizontal line
shows the background flux. From the spectrum, it is easy to find
that the microwave spectral fine structures in the form of dots
scattering in the frequency range of 0.8--0.9 GHz. The dot bursts
appear as a group of individual bursts with short duration and
narrow frequency bandwidth. The figure shows that for the single
dot burst, the time period is about several tens milliseconds and
the frequency range is several tenth MHz. Associated with the
development of the flare, these dots occurred more frequently and
brightness, and extended from low frequency side to high frequency
side. Fig.\,\ref{Fig:fig_20100801} d--f show the initial
brightening at EUV wavelengths in preflare phase and the bright
flare loops in the impulsive phase of the flare observed by
AIA/SDO.

\subsubsection{C4.4 Event on 2010 August 14}

\begin{figure}
\epsscale{1.0}
\plotone{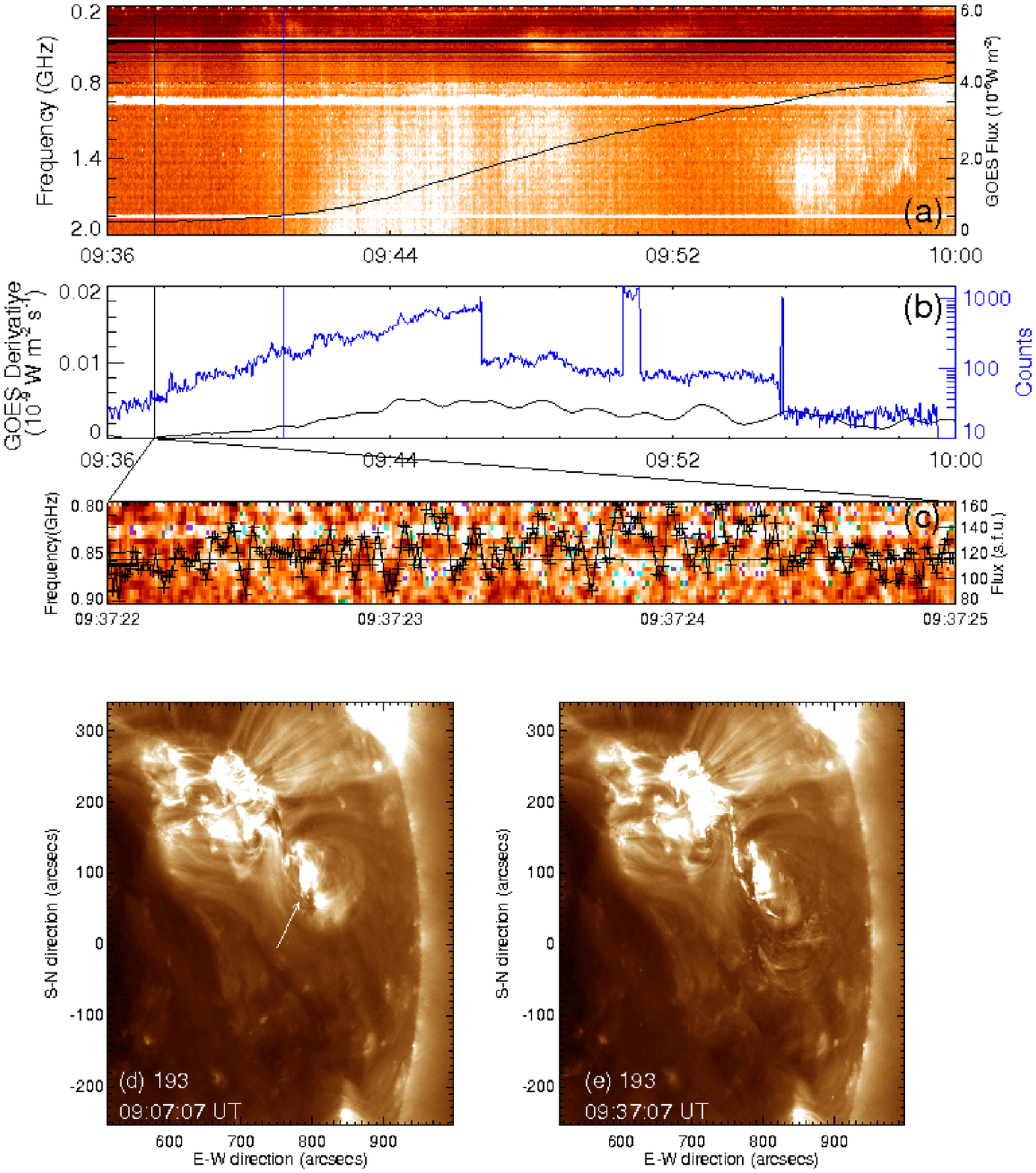}
\caption{The same figure as Fig. \ref{Fig:fig_20100801}, but for the event on 2010 August 14. }
\label{Fig:fig_20100814}
\end{figure}

The C4.4 flare event on 2010 August 14 appeared near the western
limb of solar disk(N17W52) and was associated with a filament
eruption. It is a long duration flare with impulsive phase as long
as 18 minutes. The erupted filament locates along the boundary of
two active regions as a sigmoid shape observed by EUV images in
AIA/SDO. The beginning of the activity can be marked as the slowly
rising of the filament, which occurred about forty minutes before
the GOES flare. Fig. \ref{Fig:fig_20100814}a shows the dynamic
microwave spectrum during the preflare and impulsive phase in the
frequency range from 0.2 GHz to 2.0 GHz. Black line marks a weak,
short time duration radio burst, which occurred about 4 minutes
before the flare. The GOES flux in 1--8 \AA{} energy band was
shown by the black curve. Blue line indicates the beginning of the
flare. Fig. \ref{Fig:fig_20100814}b shows the time derivative of
GOES flux in 1--8 \AA{} (black curve) and RHESSI flux at 12--25
KeV energy band (blue curve). Both curves show the slow increase
of the flux in the preflare phase. The microwave fine structure in
the preflare phase was shown in Fig. \ref{Fig:fig_20100814}c.
Radio flux at 0.87 GHz is overlapped on the spectrum and the
horizontal line shows the background flux. It shows a group of dot
bursts near frequency of 0.8 GHz. The frequency bandwidth for each
single dot is about several tenth MHz, and the duration is about
several tens of millisecond. Fig. \ref{Fig:fig_20100814}d shows
the morphology of the filament at its initial activities. And
Fig.\,\ref{Fig:fig_20100814}e shows the filament eruption during
the preflare microwave bursts.

\subsubsection{M1.9 Event on 2012 May 07}

\begin{figure}
\epsscale{1.0}
\plotone{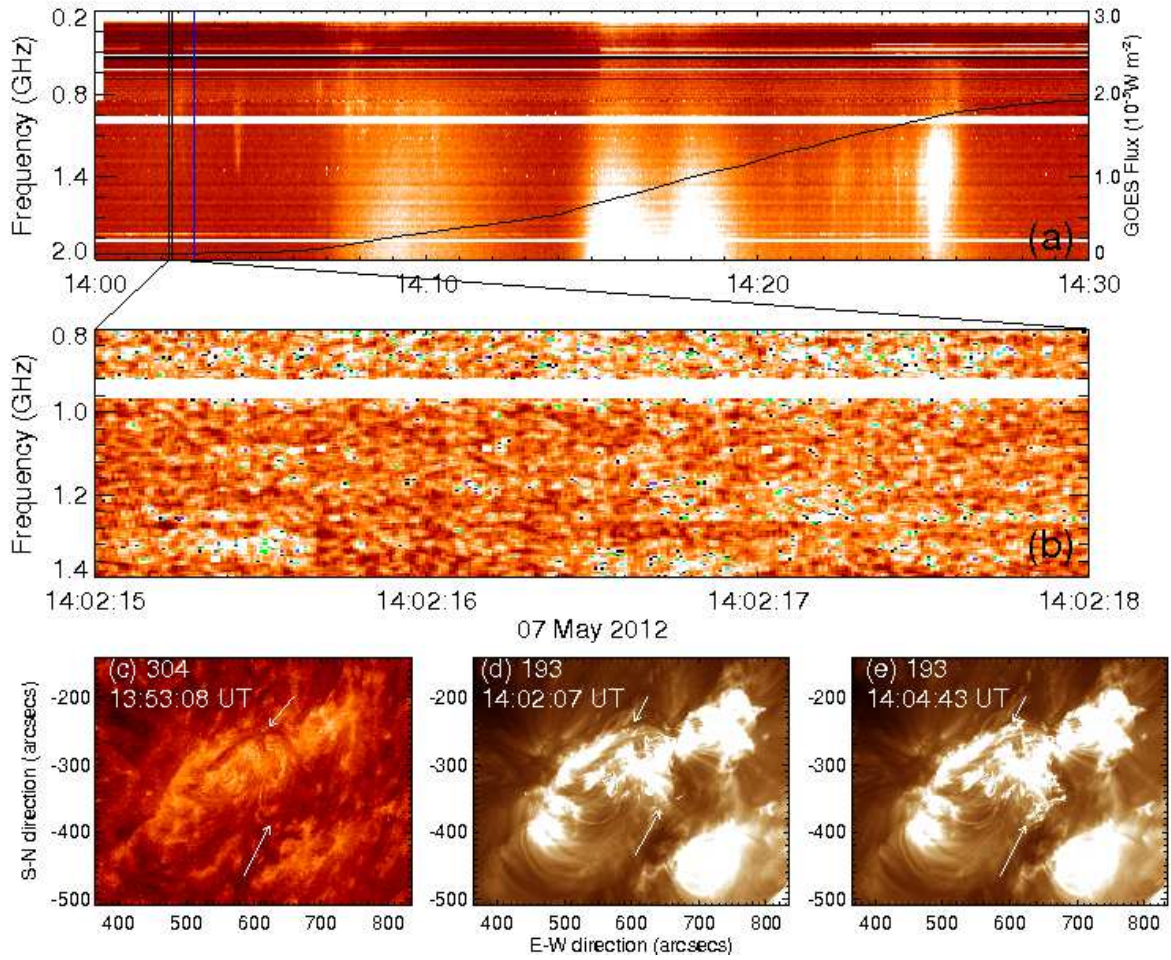}
\caption{The same figure as Fig. \ref{Fig:fig_20100801}, but for the event on 2012 May 07.}
\label{Fig:fig_20120507}
\end{figure}

The M1.9 flare event on 2012 May 7 is also a long duration event
with the impulsive phase of 28 minutes. Fig.
\ref{Fig:fig_20120507}a shows the dynamic microwave spectrum of
the flare from the very beginning to its peak. It shows that some
weak and independent radio bursts occurred just one minute before
the flare in the frequency range of 0.55--1.4 GHz. The fine
structures of the preflare microwave burst are shown by the
expanded fully temporal resolution spectrum in Fig.
\ref{Fig:fig_20120507}b. It shows that fine structures in the form
of dots crowded in a narrow band. For a single dot, the bandwidth
is about several tenth MHz, and the duration is about several tens
of millisecond. Fig. \ref{Fig:fig_20120507}c-e show that the
eruption begins as the activity of a small filament, which
locates along the margin of the host region as indicated by
arrows. The filament can not be identified in 193 \AA{} before its
eruption, while it can be identified in 193 \AA{} as some
brightening fiber with obvious twisting during the eruption.

\subsection{Brief Summary of Observational Results}

We have studied the properties of microwave fine structures in the
preflare phase and the associated activities in the source regions
observed at multiple wavelengths (such as hard X-ray of RHESSI, UV
and EUV of AIA/SDO, etc.). The results derived from the analysis
of four sampled events are summarized in Table 1. The main
observational results are as follows:

(1) There are some microwave bursts with spectral fine structures
occurring in the preflare phase of the mentioned flare events.
These fine structures appeared at about 1\,-- 4 minutes before the
start of GOES flares in the flare host region, for which direct
physical associations with flares are implied. Among the four
events, very short period QPPs occurred in one short duration
flare (C5.5 flare on 2011 December 25), while small-scale burst,
like dot burst groups, occurred in other three long duration
flares (C4.4 flare on 2010 August 01, C3.2 flare event on 2010
August 14, and M1.9 flare on 2012 May 07).

(2) There is only one type of microwave spectral fine structure
which was identified in the preflare phase for each flare. The
microwave QPPs appeared at about 2 minutes before the flare. Their
durations are in the time period from several tenth to several seconds, and
their frequency bandwidths are in the range of several hundreds MHz. Their periods are
of several tenth seconds which belong to very short period QPPs.
While the microwave dot bursts appeared at about 1--4 minutes before
the flare. They are all in great cluster with duration longer
than 10 seconds. Meanwhile, the duration and frequency range for
each dot is about several tens milliseconds and several tenth MHz.

(4) For all four events, except one occurred in the RHESSI night
time, the other three events have an obviously HXR flux
enhancement. As for the C5.5 flare on 2011 December 25,
accompanying the microwave QPPs in the preflare, the RHESSI HXR
images were obtained in 3 -- 6 keV, 6 -- 12 keV, and 12 -- 25 keV.
It shows the presence of single source in 3 -- 6 keV and 6 -- 12
keV and double sources in 12 -- 25 keV. The RHESSI sources were
located on the polarity inverse line and near the initial
brightening in UV/EUV images.

(5) From SDO observations, we found that for all events, the
beginning of the flare can be marked by filament arise or plasma
ejection, and simultaneously associated with the loop brightening
or loops interactions.

\section{Conclusions and Discussions}

This work investigated a series of solar flares and
confirmed the relationships between the preflare radio bursts and
the flare activities in four flares. From these
investigations, we find that there are several observational
activities occurring several minutes before the flare start. These
activities include:

1. Microwave bursts with spectral fine structures at
relatively weak intensity and very short timescales. These fine
structures occurred in the frequency range of 0.3 GHz -- 1.5 GHz,
including microwave QPPs with very short period of 0.1-0.3 s and
dot bursts with millisecond timescales and narrow frequency
bandwidths.

2. Filament motions, especially behaved as sigmoid filament
ascending, plasma ejections, and plasma loop brightening. These
motions can be observed by the EUV images of AIA/SDO.

3. Non-thermal processes, such as the HXR enhancement at
the energy channel of 12 -- 25 keV observed by RHESSI or the rapid
growth of the temporal derivative of GOES soft X-ray flux.

In fact, the above three different observational
activities are closely related to each other essentially. Both of
the subsecond microwave spectral fine structures and HXR
enhancements indicate the activities of electron acceleration and
primary energy release. Consequently, these releasing energy and
the energetic electrons may trigger the plasma motion in the
source region, and result in the sigmoid filament ascending,
plasma ejections, and plasma loop brightening which can be
observed by the EUV imaging observations.

Previous literatures showed that microwave bursts with
spectral fine structures frequently occurred in the impulsive and
decay phases of solar flares(Chernov 2006; Huang et al. 2008;
Huang \& Tan 2012). In this work, we find that microwave bursts
with spectral fine structures also occur in the preflare phase,
especially the microwave QPP with subseconds and dot bursts with
millisecond timescales which have similar properties as observed
in previous works. Many people have pointed out that it is very
difficult to adopt the general MHD oscillation mechanism to
explain the formation of microwave QPP with subsecond time scales
(see review of Aschwanden 1987; Nakariakov \& Milnikov 2009). \textbf{Tan
et al. (2007) proposed that the resistive tearing-mode
oscillations in current-carrying flare loops may modulate the
nonthermal microwave emission and form QPP at subsecond periods.
In this new mechanism, the plasma loops carry electric currents
with finite resistivity which may excite resistive tearing-mode
instability and produce a series of magnetic islands. Electrons
can be accelerated near the X-points between each two adjacent
islands and form energetic electron beams. The plasma may produce
coherent emission when the energetic electron beams interact with
the adjacent plasmas. When acceleration from X-points is modulated
by the tearing-mode oscillation, microwave QPP can occur (Tan et
al. 2007). The occurrence of subsecond microwave QPP implies that
the electron acceleration is quasi-periodic. When acceleration
from X-points are stochastic, microwave dot bursts may occur with
a random distribution in the spectrogram (Tan 2013). Because the
accelerating site is located around each X-point, it lasts very
short and occurs in a small region. The microwave QPPs with period of 0.1-0.3 s and dot bursts in the
preflare reveal the following information in the flare source
region: the flare loops carry electric currents, and therefore
there are considerable accumulation of non-potential free energy
which may consequently trigger the magnetic reconnection and
electron acceleration in the flare source region. Therefore,
energy release may take place in the preflare as well as in the
impulsive phase of the flares, although their intensities are
relatively weaker than the later.}

When we investigate preflare microwave signature to
understand and predict the solar flare, we always meet two
critical problems. One is the definition of the start time of a
flare, and the other is whether the preflare microwave signatures
are rare or general. In this work, the start time of flares were
extracted from GOES list. The GOES satellites provide continuous
monitoring of the integrated full disk solar X-ray intensity in a
hard (0.5 -- 4 \AA{}) channel and a soft (1 -- 8 \AA{}) channel.
The flare start defined by SXR/GOES is when four consecutive 1
minute SXR values meet all following three conditions: (1) all
four values are above the background threshold; (2) all four
values are strictly increasing; and (3) the last value is greater
than 1.4 times of the first value that occurred 3 minutes earlier.
Meanwhile, the start time is defined as the first minute in the
sequence of 4 minutes. According to this definition, it is quite
natural that some small, gradual or impulsive flux enhancement
which associated with small activities do exist before the GOES
flare.

As for the second problem, preflare activities in microwave are
discovered as early as 1950s by Bumba \& K\v{r}ivsk\'{y} (1959).
They found that about 20\% of flares in their sample are preceded
by much smaller bursts on wavelength 130 cm and 56 cm. By using
interferometric solar microwave data recorded by the Owens Valley
at 10.6 GHz between Feb 19, 1980 and March 31, 1981, Hurford \&
Zirin (1982) found that 15\% of the flares are preceded by similar
preflare signatures. The most common signature was a step-like
increase in signal amplitude, accompanied by a decrease or
reversal in the degree of polarization. Later Kai, Nakajima, and
Kosugi (1983) also found about 26\% of the flares are associated
with the preflare microwave activities which related to the
subsequent main energy release. Xie et al. (1994) studied
microwave flux variation of flares at four frequency (1.42, 2.13,
2.84, 4.26 GHz) and found that around 30\% of flares are preceded
by narrow banded preflare activities. Meanwhile, they found that
the preflare activity occurs at the low frequencies more
frequently than at the high frequencies. The rare previous studies
show that radio preflare activities do not occur in a majority of
cases. Hurford, Read, \& Zirin (1984) have shown that preflare
bursts may occur in such narrow spectral band that they can easily
be missed by instruments responding to a single frequency. Here in
the present work, we found that out of the 156 flare associated
radio bursts, 12\% (18 events) of them were preceded by microwave
fine structures. For the lack of corresponding image observations,
only 4 events are confirmed that located in the same host region
as the main flare. Such preflare activities appear to be as a
narrow-banded (not greater than 1 GHz), short duration (the
longest duration is 40 s for impulsive radio burst), and weak
intensity (about 10\% of the maximum flux of the main flare). So
they can easily be missed by narrow band spectrometers.

Preflare activities appeared in multi-wavelength have been
considered as a potential clue to understand the physical
conditions in the solar atmosphere that leads to a flare eruption.
Especially the spectral fine structures in microwave bursts, the
preflare activities can provide insights into the magnetic fields,
plasma condition, and the initial motion and interaction of the
plasma loops, and the primary magnetic reconnection process in the
flare cradle regions. Additionally, as the microwave signals are
much more sensitive than the other wavelength observations, a
systematic summary of the preflare characteristics can help us to
predict the catastrophe solar eruptive events. However, there is
no corresponding imaging observations at the related frequencies
in our study, some important parameters of the source regions,
such as the spatial scale, height, configurations of the source
region are not clear. And without the polarization information,
the parameters of the magnetic field and plasma in the source
region can not be deduced, either. These information is very
important for us to establish the direct connections of the
microwave bursts and the multi-wavelengths activities. The
forthcoming snapshot imaging observations at broad frequency
bandwidth of the next generation radio telescopes (CSRH, 0.4 -- 15
GHz, Yan et al. 2009; FASR, 0.05 -- 20 GHz, Bastian 2003) will
provide much more abundant and confirmable information about the
preflares. And then we will do further investigations on this
topic, and get more insight of the flare triggering mechanism.


\acknowledgments

The authors extend special thanks to the referee for useful
suggestions that have greatly improved the manuscript. Y. Zhang
would like to thanks Dr. Pick for her guiding in analysis NRH
data. The authors also acknowledge the AIA team for the easy
access to calibrated data. This work is supported by NSFC Grant
11373039, 11273030, 11221063, 11103044, 11433006, MOST Grant
2011CB811401, the National Major Scientific Equipment R\&D Project
ZDYZ2009-3, the Grant P209/12/00103 (GA CR), and the Marie Curie
PIRSES-GA-295272-RADIOSUN project. Y. Zhang's work is also
supported by the Young Researcher Grant of National Astronomical
Observatories, Chinese Academy of Sciences.



\begin{thebibliography}{}

\bibitem[Asai(2005)]{Asai05} Asai, A., Nakajima, H., Shimojo, M., White, S. M., Hudson, H. S., 2005, Proceedings of 9th Asian-Pacific Regional IAU meeting, 1

\bibitem[Aschwanden(1987)]{asc87}Aschwanden, M.J.: 1987, \emph{SoPh} \textbf{111}, 113

\bibitem[Bastian(1998)]{Bastian1998}Bastian, T., Benz, A.O., \& Gary, S.E.: 1998, \emph{Ann. Rev. Astron. Astrophys} \textbf{36}, 131.


\bibitem[Ben(2009)]{Ben09} Ben, A. O., Monstein, C., Meter, H., Manoharan, P. K., Ramesh, R., Altyntsev, A., et al., 2009, Earth Moon Planet, 104, 277

\bibitem[Bumba(1959)]{Bumba59} Bumba, V., \& K\v{r}ivisk\'{y}, L. 1959, Bull. Astron. Inst. Czech, 10, 221

\bibitem[Canfield(1999)]{Canfield99} Canfield, R. C., Hudson, H. S, McKenzie, D. E., 1999, Geophys Res Lett, 26, 627

\bibitem[Cheng(1985)]{cheng85} Cheng, C. C., Pallavicini, R., Acton, L. W., Tandberg-Hanssen, E., 1985, ApJ, 298, 887

\bibitem[Chernov(2006)]{chernov2006} Chernov, G. P., 2006, SSRv, 127, 195

\bibitem[Chifor(2006)]{chifor06} Chifor, C., Mason, H. E., Tripathi, D., Isobe, H., and Asai, A., 2006, A\&A, 458, 965

\bibitem[Chifor(2007)]{chifor07} Chifor, C., Tripathi, D., Mason, H. E., \& Dennis, B. R., 2007, A\&A, 472, 967

\bibitem[Contarino(2003)]{contarino03} Contarion, L., Romano, P., Yurchyshyn, V. B., Zuccarello, F., 2003, Sol. Phy, 216, 173

\bibitem[Dere(1997)]{Dere97} Dere, K. P., Landi, E., Mason, H. E., Monsignori Fossi, B. C., \& Young, P. R., 1997, A\&AS, 125,149


\bibitem[Farnik(1989)]{Farnik98} F\'{a}rn\'{\i}k, F., \& Savy, S. K., 1998, Sol. Phy, 183, 339

\bibitem[Fu(2004)]{Fu2004}Fu, Q.J., Ji, H.R., Qin, Z.H. et al., 2004a, \emph{SoPh}, \textbf{222}, 167

\bibitem[Fu(2004)]{Fu04} Fu, Q. J., Yan, Y. H., Liu, Y. Y., Wang, M., \& Wang, S. J., 2004b, CHJAA, 4, 176

\bibitem[Gaizauskas(1989)]{Gaizauskas89} Gaizauskas, V., 1989, Sol. Phy, 121, 135

\bibitem[Harrison(1985)]{Harrison85} Harrison, R. A., Waggett, P. W., Bentley, R. D., et al. 1985, Sol. Phy, 97, 387

\bibitem[Huang(2008)]{Huang08}Huang, J., Yan, Y. H., \& Liu, Y. Y., 2008, Solar Phys., 253, 143

\bibitem[Huang(2012)]{Huang12}Huang, J., \& Tan, B. L., 2012, ApJ, 745, 186

\bibitem[Hurford(1984)]{Hurford82} Hurford, G. J., Read, J. B., \& Zirin, H., 1984, Sol. Phy, 94, 413

\bibitem[Hurford(1982)]{Hurford84} Hurford, G. J., \& Zirin, H., 1982, AFGL-TR-82-0117

\bibitem[Isobe(2006)]{Isobe06} Isobe, H., \& Tripathi, D., 2006, A\&A, 449, L17

\bibitem[jiricka(1993)]{jiricka93} Ji\v{r}i\u{c}ka, M., Karlick\'{y}, M., Kepka, O., Tlamicha, A., 1993, Sol. Phy, 147, 203

\bibitem[Kai(1983)]{Kai83} Kai, K., Nakajima, H., \& Kosugi, T., 1983, Pulb. Astron. Soc. Japan, 35, 285


\bibitem[Landi(2006)]{Landi06} Landi, E., Del Zanna, G., Young, P. R., et al., 2006 ApJS, 162, 261


\bibitem[Lemen(2012)]{Lemen12} Lemen, J. R., Title, A. M., Akin, D. J., et al., 2012, Sol. Phy, 275, 17

\bibitem[Lin(2002)]{Lin02} Lin, R. P., et al., 2002, Sol. Phy, 210, 3

\bibitem[Liu(2010)]{Liu10} Liu, R., Liu, C., Wang, S., Deng, N., \& Wang, H. M., 2010, ApJ, 725, L84

\bibitem[Martin(1980)]{Martin80} Martin, S. 1980, Sol. Phy, 68, 217

\bibitem[Nakariakov and Milnikov(2009)]{Nakariakov09}Nakariakov, V.M., \& Milnikov, V.F.: 2009, \emph{Space Sci. Rev.} \textbf{149}, 119.

\bibitem[Priest(2002)]{Priest02} Priest, E. R., \& Forbes, T. G., 2002, A\&ARv, 10, 313

\bibitem[Shibata(1999)]{Shibata99} Shibata, K., 1999, Ap\&SS, 264, 129

\bibitem[Tan(2007)]{Tan07}Tan, B.L., Yan, Y.H., Tan, C.M., \& Liu, Y.Y.: 2007, ApJ, 671, 964

\bibitem[Tan(2010)]{Tan2010}Tan, B. L., Zhang, Y., Tan, C. M., \& Liu, Y. Y., 2010, ApJ, 723, 25

\bibitem[Tan(2012)]{Tan2012}Tan, B.L., Tan, C.M., 2012, ApJ, 749, 28

\bibitem[Tan(2013)]{Tan2013}Tan, B.L., 2013, ApJ, 773, 165

\bibitem[Tappin(1991)]{Tappin91} Tappin, S. J., 1991, A\&As, 87, 2Tan, C.M.77

\bibitem[Vemareddy(2012)]{Vemareddy12} Vemareddy, P., Maurya, R. A., \& Ambastha, A., Sol. Phy, 277, 337

\bibitem[warren (2001)]{warren01} Warren, H. P, Warshall, A. D., 2001, ApJ, 560, L87

\bibitem[xie(1994)]{xie94} Xie, R. X., Song, Q., Wang, M., Chen, G. Q., 1994, Sol. Phy, 155, 113

\bibitem[Yan et al(2009)]{Yan09}Yan, Y.H., Zhang, J., \& Wang, W., et al. 2009, \emph{EM$\&$P} \textbf{104}, 97.

\end{thebibliography}
\end{document}